\documentstyle[12pt,]{article}
\begin{document}
\centerline{\bf{COSMIC EVOLUTION IN }}
\centerline{\bf{GENERALISED BRANS-DICKE THEORY}}
\vspace{0.2in}
\centerline{B.K.Sahoo$^{1}$ and L.P.Singh$^{2}$}
\centerline{Department of Physics,}
\centerline{Utkal University,Bhubaneswar-751004,India}
\centerline{$^{1}$  bijaya@iopb.res.in}
\centerline{$^{2}$  lambodar@iopb.res.in}
\vspace{0.25in}
\begin{abstract}
We have studied the Generalised Brans-Dicke theory
and obtained exact solutions of a(t),$\phi(t)$, and $\omega(t)$ for
different epochs of the cosmic evolution. We discuss
how inflation,decceleration,cosmic acceleration can 
result from  this solution.The time variation of G(t) is also
examined.
\end{abstract}
PACS NO:98.80.-K,98.80.cq
\centerline{accelerated expansion,inflation,time variation of G(t).}
\newpage
\section{INTRODUCTION}
The Brans-Dicke Theory(BD) is defined by a scalar field 
$\phi$ and a constant coupling function $\omega$[1].This is
 perhaps the most natural extension of general theory of  relativity (GR) which
 is obtained in the limit of $\omega\rightarrow\infty$ and $\phi$=
constant[2].This theory interestingly appears naturally in 
supergravity theory,Kaluza-Klein theories and in all the
 known effective string actions.One important property of BD theory
is that it gives simple expanding solutions[ 
     3]for field $\phi(t)$
 and scale factor $a(t)$ which are compatible with solar system 
experiments[4].The Generalised Brans-Dicke
theory(GBD)[2],sometimes referred to as graviton-dilaton or scalar
-tensor(ST) theory, is instead, defined by $\omega$ which is an
arbitrary function of the
scalar field $\phi$ (dilaton) and is  thus implicitly a  function
of time $\omega(\phi(t))$.Hence,it includes a number of
 models,for every function $\omega$.Naturally,a few attempts 
have been taken to study the dynamics of the universe using 
this formalism.Some of the cosmological problems studied
 under this formalism includes nonsingular universe[5],amplification
 of gravitational wave[6],solutions for different epochs of cosmic
 evolution and time variation of Newton's gravitational constant G[7]
etc.These problems are studied for a specific forms of $\omega(\phi)$ 
involving quite  complicated time-dependence for a(t),$\phi(t)$.Thus,it is
difficult to draw a clear physical  picture about the evolution of the
universe.\\
 \indent  In order to carry out a detailed study of the dynamics of the 
cosmic evolution in this formalism,knowledge about exact
time-dependence of a(t),$\phi(t)$ and $\omega(t)$ for different epochs 
is essential.
            In a previous work[8],we have obtained power law
 time-dependence of $\omega(t)$ with the power of time determined in
 terms of the exponent of  scale factor a(t) which is
 taken to vary as $a(t)=a_{0}\left(\frac{t}{t_{o}}\right)^{\alpha}$.To fix the
 value $\alpha$ we had taken the help of observational
 information.Similar result is also arrived at by Diaz-Rivera and Pimentel
 [9].However,in these works the value of $\alpha$ can not be 
fixed purely by way of solving the gravitational and scalar field equations.
In the present work we take our previous attempt a little further and
explore  to see if a consistent
solution of the GBD equations of motion based on simple power-law
temporal behaviour of a,$\phi$ and $\omega$ can fix the time exponents.\\
\indent We have arranged this paper in the following manner.In 
the section 2 we obtain the solutions for a,$\phi$,$\omega$ for 
different epochs.We present a discussion about the present value of 
$\omega$ in section 3 and examined the time 
variation of Newton's gravitational constant G in  section 4 and
conclude with section 5.
 \section{FIELD EQUATIONS AND SOLUTIONS}
  For a universe filled with perfect fluid and described by Friedman-
Robertson-Walker space-time with scale factor $a(t)$ with spatial
curvature index k,the gravitational field equations in GBD theory are
\begin{equation}
\frac{\dot{a}^{2}+k}{a^{2}}+\frac{\dot{a}\dot{\phi}}{a\phi}-\frac{\omega
\dot{\phi}^{2}}{6\phi^{2}}=\frac{\rho}{3\phi}.
\end{equation}

\begin{equation}2\frac{\ddot{a}}{a}+\frac{\dot{a}^{2}+k}{a^{2}}+\frac{\omega{\dot{\phi}^{2}}}{2\phi^{2}}+2\frac{{\dot{a}}{\dot{\phi}}}{a\phi}+\frac{{\ddot{\phi}}}{\phi}=-\frac{P}{\phi}.
\end{equation}
where $\rho$ and P are respectively the energy density and pressure of 
the  fluid distribution.The equation of state of the fluid is given by
$P=\gamma\rho$.Some of the values of $\gamma$ for typical cases are -1
(vacuum),0(dust),1/3(radiation),1(massless scalar field).
          The wave equation for scalar field in the GBD  theory is [2] 

\begin{equation}
\ddot{\phi}+3\frac{\dot{a}\dot{\phi}}{a}=\frac{\rho-3P}{2\omega+3}-\frac{\dot{\omega}\dot{\phi}}{2\omega+3}.
\end{equation}
Energy conservation equation, which can be obtained from
eqs.$(1)$,$(2)$ and $(3)$ is,

\begin{equation}
\dot{\rho}+3\frac{\dot{a}}{a}\left(\rho+P\right)=0.
\end{equation}
To obtain a cosistent
solution for the equations of motion we assume simple power law
dependence on time for a,$\phi$ and $\omega$ in the following form
which provides simple expanding solutions[6-10],
\begin{equation}
a=a_{0}\left(\frac{t}{t_{0}}\right)^{\alpha}.
\end{equation}

\begin{equation}
\phi=\phi_{0}\left(\frac{t}{t_{0}}\right)^{\beta}.
\end{equation}

\begin{equation}
\omega(\phi(t))\approx\phi^{n}=\omega_{0}\left(\frac{t}{t_{0}}\right)^{n\beta}.
\end{equation}
$\alpha$ is a positive number to ensure an  expanding  universe and
$n$ is a non zero constant.$\alpha$,$\beta$ and $n$ are
the free parametres at this level whose values we wish to fix by way
of obtaining consistent solutions of the gravitational and scalar
field equations of motion.$a_{0}$, $\phi_{0}$ are present values of
a(t) and $\phi(t)$.We further assume  $k=0$ to remain consistent
with inflationary paradigm,which has received strong observational
support.Before proceeding further,it may be noted that eqn(4) has 
the solution,\\ 
\centerline{ $\rho=C a^{-3(1+\gamma)}$}\\
and so
\begin{equation}
\rho=\rho_{0}\left(\frac{t}{t_{0}}\right)^{-3\alpha\left(1+\gamma\right)}
\end{equation}
$\rho_{0}=C a_{0}$is the  present($t=t_{0}$) value of energy density.\\
We now discuss the solutions of the field equations in different
epochs.

\subsection{Radiation dominated epoch: $\gamma=1/3$}

For radiation dominated  epoch the gravitational field equation [eqn.(1)]and
 scalar field wave equation [eqn.(3)] for  k=0 become
\begin{equation}
\left( \frac{ \dot{a}} {a} \right)^{2} = \frac { \rho}{3\phi} - \frac{ 
  \dot{a} \dot{\phi} } {a \phi} + \frac{\omega}{6} \left( \frac
  {\dot{\phi}}{\phi}\right)^{2}.
\end{equation}
and \begin{equation}
\ddot{\phi} + 3 \frac{ \dot{a}}{a} \dot{\phi} + \frac{ \dot{\omega}
  \dot{\phi}} {2\omega + 3} = 0 .
\end{equation}
\vspace{.1in}
\noindent{Using equation (5),(6),(7) and (8), eqn.(9) yields}\\
\vspace{.1in}
\centerline{ $\left( \frac{ \alpha}{t}\right)^{2}    =  
\frac{\rho_{0}}{3\phi_{0}} \left(\frac{t}{t_{0}}\right)^{-4\alpha-\beta}
- \frac{\alpha \beta}{t^{2}} + \frac{\omega_{0}}{6} \left( \frac
{t}{t_{0}}\right)^{ n \beta } \frac{ \beta^{2}}{t^{2}}$} \\
\vspace{.1in}
\noindent{This equation can be put in the form,}\\
\vspace{.1in} 
\centerline{$\left( \alpha^{2} + \alpha \beta \right) - 
\frac{ \rho_{0} t_{0}^{2}} { 3\phi_{0}} \left( \frac{t}{t_{0}}\right)
^{ -4\alpha -\beta + 2} = 
\frac{ \omega_{0} \beta^{2} }{6} \left( \frac{t}{t_{0}} \right)^{ n
  \beta }$}
This equation is satisfied when\\
\begin{equation}
-4 \alpha - \beta + 2 = n \beta 
\end{equation}
\begin{equation}
\alpha^{2} + \alpha \beta = 0 
\end{equation}
\begin{equation}
\beta^{2} = - \frac{2 \rho_{0} t_{0}^{2}}{\omega_{0} \phi_{0}} .
\end{equation}
Eqn.(10), in similar manner leads to
\begin{equation}  \beta = 0 
\end{equation}
and
\begin{equation}
\beta = \frac { (1-3\alpha)(2 \omega_{0} + 3 )} { (n + 2
  )\omega_{0} + 3 }
\end{equation}
\vspace{.1in}
Here we have further assumed \\
\centerline{
$\Omega(t)=2\omega(t)+3
=\Omega_{0}\left(\frac{t}{t_{0}}\right)^{n\beta}$,
where $\Omega_{0} = 2\omega_{0} +3 $.}
Power law solutions  (5),(6),(7) satisfy  eqns.(9)and (10)
simultaneously when  eqn.(11)and  eqn.(12) are simultaneously
satisfied with $\beta$ given by either eqn,(14) or eqn.(15).\\
\indent The $\beta=0$ solution of eqn.(14) implies  $\alpha=1/2$ (from
eqn.(11)) and $\phi$=constant and $\omega$=constant which represents 
the well known Standard Model solutions.Thus, $\beta = 0 $ solution 
represents the  GR sector of  GBD theory.\\
\vspace{.1in}
Now using  eqn.(15) in eqn.(11)one  gets\\
\vspace{.1in}
\centerline{$ \alpha = \frac{ 3 n - 2 \omega_{0} - 3 } {2 n \omega_{0} - 2
  \omega_{0} + 9 n -3 }$}\\
\noindent{Similarly,eqn.(15)in conjunction with  eqn.(12) gives}\\
\begin{equation}\alpha =  \frac{ 2\omega_{0} +3 } { (4- n )\omega_{0}
    + 6 }   
\end{equation}
For eqns(11),(12) and (15) to be simultaneously satisfied ,
above two values of  $\alpha$  should be
equal.
\vspace{.1in}
Imposition of this condition leads to \\
\vspace{.1in}
\centerline{$3\omega_{0} n ^{2} + (2\omega_{0}^{2} + 9 \omega_{0} + 9 ) n + 
(6 \omega_{0}^{2} + 15 \omega_{0} + 9 ) = 0 $}\\
which  admits two solutions like,
$ n = -\frac{(2\omega_{0} + 3)} {\omega_{0}} $  and   $- \frac{
    (2\omega_{0} + 3 )}{3} $.\\
Substitution of these two values of  n  in eqn.(17),leads to 
   $\alpha = 1/3 $  and $\alpha=\frac{3}{\omega_{0}+6}$ respectively.
Once again we find from eqn.(15)for $\alpha=1/3$, $\beta$ is  equal to
zero implying
constant $\phi$ and constant $\omega$ .We neglect this solution as it
does not capture the characteristics of BD theory.Now for the second
solution $\alpha=\frac{3}{\omega_{0}+6}$ we find  
 $\beta  = -3/(\omega_{0}+6)$ and  $n\beta = \frac{2\omega_{0} +
  3}{\omega_{0}+6}$.
Hence, the time dependences  are obtained as
\begin{equation}
 a(t)= a_{0}\left(
  \frac{t}{t_{0}}\right)^{\frac{3}{\omega_{0}+6}}
\end{equation}

\centerline{$\phi(t)=\phi_{0}\left( \frac{t}{t_{0}}
\right) ^{-\frac{3}{\omega_{0}+6}}$}

 \centerline{$\omega(t)= \omega_{0}\left(\frac{t}
{t_{0}}\right)^{ \frac{2\omega_{0} + 3}{ \omega_{0}+6}}$}

As can be seen from above ,the time dependence of  $a(t)$
$\phi(t)$and $\omega(t)$ are completely fixed once $\omega_{0}$ 
value is known.$\omega_{0}$ needs to be fixed from other physical 
considerations.One can recover the
 standard model result for $a(t)$ from above equations when
 $\omega_{0}=0$.Further it is easily seen that for ensuring
 deccelerated expansion in the radiation epoch($0 < \alpha < 1$)
 $\omega_{0} $ should be greater than  $-3$.The values for $\beta$ 
and $n\beta$ are obtained to lie in the range $-1< \beta <0$ and 
$-1< n\beta <2$ respectively.
\subsection{Matter dominated epoch: $\gamma=0$ }
 
Using eqn.(5),(6) and(7), in eqn(1) we get\\

$\left( \frac{\alpha}{t}\right)^{2} = 
1/3 \frac{1}{\phi_{0}} \left( \frac{t}{t_{0}}\right)^{-\beta} \rho_{0}
\left(\frac{t}{t_{0}}\right)^{-3\alpha} - \frac{\alpha \beta}{t^{2}} + 
\frac{\omega_{0}}{6} \left( \frac{t}{t_{0}}\right)^{n \beta} \left(
  \frac{\beta}{t}\right)^{2}$,\\

\noindent{which can be written in the form}\\

$- \frac{ \rho_{0} t_{0}^{2}}{3 \phi_{0}} \left(\frac{t}{t_{0}}\right)
^{2-3\alpha-\beta} + \left( \alpha^{2} + \alpha \beta\right) = 
\frac{\beta^{2} \omega_{0}}{6} \left( \frac{t}{t_{0}}\right)^{ n \beta 
  }$.\\

\noindent{This equation is satisfied when}\\
\begin{equation}
2- 3\alpha - \beta = n \beta
\end{equation}
\begin{equation}
\alpha^{2} + \alpha \beta = 0
\end{equation}
\begin{equation}
\beta^{2}=-2\frac{\rho_{0}t_{0}^{2}}{\phi_{0}\omega_{0}}
\end{equation}
\vspace{.1in}
Similarly,using eqn.(5),(6),(7) in eqn.(3) one gets\\
$\left(\beta(\beta-1) + 3\alpha \beta + \frac{ n \beta^{2}\omega_{0}}{\Omega_{0}}\right)\frac{\phi_{0}}{t_{0}^{2}}\left(\frac{t}{t_{0}}\right)^{\beta - 2} = 
\frac{\rho_{0}}{\Omega_{0}} \left( \frac{t}{t_{0}}\right)^{-3
\alpha - n \beta}$.\\

\noindent{This equation is satisfied when}\\

\begin{equation}
\beta-2=-3\alpha-n\beta
\end{equation}
and\begin{equation}
\left(\beta(\beta-1) + 3\alpha \beta + \frac{n \beta^{2}\omega_{0}}
{\Omega_{0}}\right)\frac{\phi_{0}}{t_{0}^{2}} = \frac{\rho_{0}}{\Omega_{0}},
\end{equation}\\
With the help of eqn.(20),eqn.(22) reduces to\\

$\left( 1+ \frac{ n \omega_{0}}{2\omega_{0}+3} + \frac{ \omega_{0}}{2(
2\omega_{0}+3)} \right) \beta^{2} + (3\alpha-1) \beta = 0 $.\\

\noindent{The above equation  leads to,$\beta=0$ }\\
and \begin{equation}
  \beta = \frac{(1-3\alpha)(4\omega_{0}
+ 6)}{(5+ 2n)\omega_{0} + 6}.
\end{equation}\\
Inserting $\beta=0$,eqn.(21) leads to $\alpha=2/3$.This is the  standard
model result representing GR sector of GBD theory.
\vspace{.1in}
Using eqn.(23), eqn.(19) leads to 
\centerline{$\alpha=\frac{4\omega_{0}+6}{(7-2n)\omega_{0}+12}$}.\\
\vspace{.1in}
\noindent{Also for eqn.(23),eqn.(18) leads to }\\
\vspace{.1in}
\centerline{$\alpha = \frac{(n+1)(4\omega_{0}+6)-2[5\omega_{0}+2\omega_{0}n+6]}
{3[(n+1)(4\omega_{0}+6)-(5\omega_{0}+2\omega_{0} n + 6)]}$}.\\
\noindent{For consistent solution of field equations,these two values of  $\alpha$ 
 should be equal}.And we find\\
\vspace{.1in}
\centerline{${2\omega_{0} n^{2}+[2\omega_{0}^{2}+9\omega_{0}+6]+[5\omega_{0}^{2}+
16\omega_{0}+12]=0}$.}\\
\noindent{Which admits two solutions
like,$n=-\frac{(5\omega_{0}+6)}{2\omega_{0}}$
 and ,$n = -(\omega_{0}+2)$}.\\
The first value of n is neglected as it once again  gives
$\alpha=\frac{1}{3}$ and $\beta=0$ as in radiation dominated case.
We ignore this solutions since it does not capture the time-varying 
characteristic of $\phi(t)$ and $\omega(t)$ specific to the GBD
theory.\\
\vspace{.1in}
\noindent The second value of n leads to  $\alpha=\frac{2}{\omega_{0}+4}$.\\ 
Hence,the relevant solutions in the matter-dominated epoch are,\\
\begin{equation}a(t)
  =a_{0}(\frac{t}{t_{0}})^{\frac{2}{\omega_{0}+4}}
\end{equation}\\
\centerline{$\phi(t)=\phi_{0}\left(\frac{t}{t_{0}}\right)^{-\frac{2}{\omega_{0}+4}}$}\\
\centerline{$\omega(t) = \omega_{0} \left(\frac{t}{t_{0}}\right)^{\frac
{2(\omega_{0}+2)}{\omega_{0}+4}}$}\\
Once again it is found that the time-dependences of $a(t)$,$\phi(t)$,
and $\omega(t)$ are  entirely controlled by $\omega _{0}$,the present value of 
$\omega $ as was also the case in the radiation-dominated epoch.For
deccelerated cosmic expansion $\omega_{0}$ need be greater than $-2$.
That way even very large values of $\omega_{0}$ required for agreement 
with solar system measurements[2] can also accomodated.The presently
observed accelerated expansion too can be accomodated if
$-4<\omega_{0}<-2$.The values of $\beta$ and $n\beta$ are found to lie 
in the range $-1<\beta<0$ and $0<n\beta<2$ respectively.
\subsection{Vacuum dominated epoch:$\gamma=-1$}
\vspace{.1in}
Using eqn.(5),(6) and (7),eqn.(1) reduces to\\

\centerline{$\left(\frac{\alpha}{t}\right)^{2} = 
\frac{\rho_{0}}{3\phi_{0}}
\left(\frac{t}{t_{0}}\right)^{-\beta}-\frac{\alpha
\beta}{t^{2}} +
\frac{\omega_{0}}{6}\left(\frac{t}{t_{0}}\right)^
{n \beta}
\left(\frac{\beta}{t}\right)^{2}$}.\\
This can be written as\\
\centerline{$\left(\alpha^{2}+\alpha\beta\right)- \frac{\beta^{2}\omega_{0}}{6}
\left(\frac{t}{t_{0}}\right)^{n \beta} =
\frac{\rho_{0}t_{0}^{2}}{3\phi_{0}}\left(\frac{t}{t_{0}}\right)^{2-\beta}$}\\
This equation is satisfied when\\
\begin{equation}
2-\beta = n \beta
\end{equation}
\begin{equation}
\alpha^{2}+\alpha\beta = 0
\end{equation}
\begin{equation}
-\frac{\beta^{2}\omega_{0}}{6} = \frac{\rho_{0}t_{0}^{2}}{3\phi_{0}}
\end{equation}
\vspace{.1in}
Similarly,using eqn.(5),(6),(7),eqn.(3) reduces to\\
\vspace{.1in}
\centerline{$\left(\beta(\beta-1)+3\alpha\beta + \frac{n \beta^{2}\omega_{0}}
{\Omega_{0}}\right)\frac{\phi_{0}}{t_{0}^{2}} \left(\frac{t}{t_{0}}
\right)^{\beta-2} = \frac{4\rho_{0}}{\Omega_{0}} \left(\frac{t}
{t_{0}}\right)^{-n\beta}$}\\
This equation is satisfied when\\
\begin{equation}
2-\beta=n \beta
\end{equation}
and
\centerline{$\beta(\beta-1) + 3\alpha\beta + \frac{n
  \beta^{2}\omega_{0}}{\Omega_{0}}
=\frac{4\rho_{0} t_{0}^{2}}{(2\omega_{0}+3)\phi_{0}}$}\\
The above equation with the help eqn.(27) reduces to 
\begin{equation}
\left(1+\frac{(n+2)\omega_{0}}{(2\omega_{0}+3)}\right)
\beta^{2}+\left(3\alpha-1\right)\beta = 0
\end{equation}
The $\beta=0$ solution of this equation which leads to inconsistency
with respect to eqns(25) and (28),we believe,is indicative of the
impossibility of obtaining exponential standard model  expansion
within the power law assumptions considered in this work.
Continuing further,equation(28) and (29) give
\centerline{$\alpha = \frac{3(n-1)-6\omega_{0}}{3(2\omega_{0}+3)(n+1)}$}\\
However,equation(25) and (26)give\\
\begin{equation}
\alpha = -\frac{2}{n+1}
\end{equation}
Now these two values of $\alpha$ are equated for consistent solutions
of equation(1) and (3).Hence,we get\\
\begin{equation}
n = -(2\omega_{0}+5)
\end{equation}
\vspace{.1in}
This value of  n leads to
$\alpha = \frac{1}{\omega_{0}+2}$
and $ \beta =-\frac{1}{\omega_{0}+2}$\\
Hence, the exact time dependence of a(t),$\phi(t)$ and $\omega(t)$ in
case of vacuum energy dominated epoch are obtained as,\\
\begin{equation}
a(t) =a_{0}\left(\frac{t}{t_{0}}\right)^{\frac{1}{\omega_{0}+2}}
\end{equation}
\centerline{$\phi(t) =
\phi_{0}\left(\frac{t}{t_{0}}\right)^{-\frac{1}{\omega_{0}+2}}$}\\
\vspace{.1in}
\centerline{$\omega(t)=
\omega_{0}\left(\frac{t}{t_{0}}\right)^{\frac{2\omega_{0}+5}{\omega_{0} 
+2}}$}\\
Once again, as in earlier two cases discussed, the time dependence is
controlled by $\omega_{0}$,the present value of  $\omega$.For
accelerated cosmic expansion we must have $-2<\omega_{0}<-1$. 
In fact,closer the value of $\omega$ to $-2$ faster is the accleration
which can mimic inflation.Such a power-law inflation was indeed
obtained by Mathiazhagan and Zohri,and La and Steinhardt[3] in their
pioneering works.Further in this epoch we obtain $\beta$ and $n\beta$
values to lie in the range $-\infty<\beta<-1$ and $3<n\beta<\infty$ 
respectively for above stated range of $\omega_{0}$.
\section{Present value of $\omega(\phi)$}
In order to look for answer to the recently observed accelerated
expansion  of the universe,structure formation,coincidence problem
etc.some authers have studied cosmological models in Brans-Dicke
theory or some modified version of this theory[10].A small negative
value of $\omega$ at  t $\rightarrow \infty $ lying in the range 
$-2<\omega_{0}<-3/2$ has been a recurrent result in these studies
,in contrdiction with large values of the order of 600[2] obtained 
from solar system measurements. A possible solution of this
contradiction,we believe, may lie in considering a non-constant 
coupling function
$\omega(\phi)$ as in GBD theory.Thus, the value of such a function can
change with the cosmic time and,in the limit  $t \rightarrow \infty$,
it could agree with local measured values.\\
\indent Further,as evident from our solutions,the time dependence
of $ a(t),\phi(t) $ and $\omega(\phi(t))$ in various epochs are
entirly controlled by $\omega_{0}$,the present value of $\omega$ 
and not by the values of $\omega$ during the epochs considered.
This brings out the importance of the value of $\omega_{0}$.
For our results obtained in this work to be consistent with the 
results obtained by earlier authers[10]
,$\omega_{0}$ can  safely  lie in the range $-2<\omega_{0}<-1$
excluding the $\omega_{0}=-3/2$ value for which BD theory 
breaks down.Thus,we arrive at a comprehensive picture of cosmic 
evolution in the important epochs with the possibility of obtaining
sufficient inflation in vacuum dominated era as well as accelerated
expansion in the present epoch.Given the simple assumptions considered 
by us, we consider this a very interesting result which signifies the
power of a GBD theory.
\section{VARIATION OF NEWTON'S GRAVITATIONAL CONSTANT G:}
As is well known, for theories with constant $\omega$,the possibility
of variations of G is very small.Consideration of arbitrary coupling
function  $\omega(\phi)$,however,opens the possibility of
variations of G.In the weak field limit Nordtvedt[2] found an
expression for the observed value of the gravitational constant\\
\begin{equation}
G(t)=\left( \frac{4+2\omega(\phi)}{3+2\omega(\phi)}\right) \phi^{-1}
\end{equation}
The variation rate of G is[2]\\
\begin{equation}
\frac{\dot{G}}{G}=-[\frac{\dot{\phi}}{\phi}+\frac{2\dot{\omega}}{
(3+2\omega)(4+2\omega)}]
\end{equation}
where over dots represent time derivative.\\
The present variation rate,therefore,is given by \\
\begin{equation}
\left(\frac{\dot{G}}{G}\right)_{0} =
-\left(\frac{\dot{\phi}}{\phi}\right)_{0}-
\frac{2\left(\dot{\omega}\right)_{0}}{(3+2\omega_{0})(4+2\omega_{0}
)}
\end{equation}
For  typical values of $\omega_{0}$like $-1.9$ or $-1$,we obtain from
the above expression,\\
$\left|\frac{\dot{G}}{G}\right|_{0}\approx 1.3 H_{0} \approx 2\times10^{-10}$
$yr^{-1}$. 
Thus,the variation of G  is safely  below the observational limit,which is $4\times 
10^{-10} yr^{-1}$[10,11] obtained within the present formalism.
\section{CONCLUSIONS}
In this work,we have studied a homogeneous and isotropic cosmological
model in the context of a generalised scalar-tensor theory. We have
solved consistently both the gravitational field equation and scalar
field  wave
equation for matter,radiation and vacuum dominated epochs of the
cosmic evolution by assuming power law time-dependence for the 
scale factor,scalar field and coupling parameter
 as extension of our earlier work[8].The solutions are generic which
 deserves attention.For all epochs,the solutions depend on one
 parameter i.e,on the present value of $\omega$.Interestingly,general 
considerations about cosmic evolution can constraint this value to lie 
in the range $-2<\omega_{0}<-1$.This result is quite consistent with other 
results[10] obtained from
various other considerations.Also,we have shown that, when
  $\omega_{0}$ is chosen very close to -2 one can produce enough
inflation according to wish,which can solve inflationary problems 
very easily in the standard model.While obtaining usual decceleration
when the universe is radiation or matter dominated we also exhibit the 
possibility of accelerated expansion in matter dominated era for a
slightly  different value of $\omega_{0}$.\\
\indent We also find that the B-D scaler field $\phi(t)$ is always a
decreasing function of time in all epochs with the decrease becoming 
very significant in vacuum dominated case.The time dependence of $\omega(t)$
,on the contrary,does not exibit such monotonic temporal behaviour.\\
\indent We have also examined the time variation
of G(t)in this solution and found that the solution gives a safe time 
variation for the chosen range of $\omega_{0}$ which gives a strong
support to the results presented here.\\

\centerline{Acknowldgement}
The authors are grateful to DST ,Govt.of India for providing financial 
support.The authors also thank Institute of Physics,Bhubaneswar,India,
for providing facility of the computer center.
 
\end{document}